\documentclass[12pt, letterpaper]{article} 
\makeatletter\if@twocolumn\PassOptionsToPackage{switch}{lineno}\else\fi\makeatother
\usepackage[margin=1in,footskip=0.25in]{geometry}

\usepackage{tabulary,graphicx,amsmath,amsfonts,amssymb}
\usepackage[utf8x]{inputenc}
\usepackage[T1]{fontenc}
\usepackage{cmbright}
\usepackage{setspace}
\usepackage[labelfont=bf]{caption}

\usepackage{authblk}
\title{Microscopic-scale recording of brain neuronal electrical activity using a diamond quantum sensor}
\author[1,a]{Nikolaj Winther Hansen}
\affil[1]{Department of Neuroscience, University of Copenhagen, 2200 Copenhagen, Denmark}
\author[2,*,a]{James Luke Webb\thanks{jaluwe@fysik.dtu.dk}}
\affil[2]{Center for Macroscopic Quantum States (bigQ), Department of Physics, Technical University of Denmark, 2800 Kgs. Lyngby, Denmark}
\author[2,a]{Luca Troise}
\author[3]{Christoffer Olsson}
\affil[3]{Department of Health Technology, Technical University of Denmark, 2800 Kgs. Lyngby, Denmark}
\author[4]{Leo Tomasevic}
\affil[4]{Danish Research Center for Magnetic Resonance, Center for Functional and Diagnostic Imaging and Research, Copenhagen University Hospital - Amager and Hvidovre, 2650 Hvidovre, Denmark}
\author[5]{Ovidiu Brinza}
\affil[5]{Laboratoire des Sciences des Proc\'ed\'es et des Mat\'eriaux, Universit\'e Sorbonne Paris Nord,  93430 Villetaneuse, France}
\author[5]{Jocelyn Achard}
\author[6]{Robert Staacke}
\affil[6]{Division Applied Quantum System, Felix Bloch Institute for Solid State Physics, Leipzig University, 04103, Leipzig, Germany}
\author[6]{Michael Kieschnick}
\author[6]{Jan Meijer}
\author[3,4]{Axel Thielscher}
\author[4,5,6]{Hartwig Roman Siebner}
\affil[5]{Department of Neurology, Copenhagen University Hospital Bispebjerg and Frederiksberg, 2400 Copenhagen, Denmark}
\affil[6]{Department of Clinical Medicine, Faculty of Health and Medical Sciences, University of Copenhagen, 2200 Copenhagen N, Denmark}
\author[3]{Kirstine Berg-S{\o}rensen}
\author[2]{Jean-Fran\c{c}ois Perrier}
\author[2]{Alexander Huck\thanks{alexander.huck@fysik.dtu.dk} }
\author[2]{Ulrik Lund Andersen\thanks{ulrik.andersen@fysik.dtu.dk}}

\affil[a]{Authors contributed equally to this work}

\date{}                     
\begin{document}

\onehalfspacing
\pagenumbering{gobble}
\maketitle
\clearpage

\begin{abstract}
An important tool in the investigation of the early stages of neurodegenerative disease is the study of dissected living tissue from the brain of an animal model. Such investigations allow the physical structure of individual neurons and neural circuits to be probed \cite{Wilt2009,Streich2021} alongside neuronal electrical activity, disruption of which can shed light on the mechanisms of emergence of disease. Existing techniques for recording activity rely on potentially damaging direct interaction with the sample, either mechanically as point electrical probes or via intense focused laser light combined with highly specific genetic modification and/or potentially toxic fluorescent dyes\cite{Scanziani2009}. In this work, we instead perform passive, microscopic-scale recording of electrical activity using a biocompatible quantum sensor based on colour centres in diamond \cite{Taylor2008}. We record biomagnetic field induced by ionic currents in mouse \textit{corpus callosum} axons without direct sample interaction, accurately recovering signals corresponding to action potential propagation while demonstrating \textit{in situ} pharmacology during biomagnetic recording through tetrodotoxin inhibition of voltage gated sodium channels. Our results open a promising new avenue for the microscopic recording of neuronal signals, offering the prospect of high resolution imaging of electrical circuits in the living mammalian brain.
\end{abstract}

\section{Main}

Neurodegenerative diseases in the brain are characterised by substantial microscopic-scale structural changes, including amyloid-$\beta$ (A$\beta$) and tau tangles aggregation \cite{Sasaguri2017, Romoli2021} or inflammation in the case of multiple sclerosis \cite{Compston2008}, which can occur well before disease becomes symptomatic \cite{Jack2009, Villemagne2013}. This includes microscopic physical structural changes associated with alterations of action potential propagation \cite{Romoli2021, Schauf1974}. Studying small-scale functional changes in the very early stages of degeneration is of utmost importance in understanding the mechanisms of disease emergence and for the eventual development of early intervention treatment. A key tool towards this goal is the study of the early onset of disease induced in an animal model \cite{Dawson2018,Sasaguri2017,Ransohoff2012}. Such studies of early onset are currently achieved with methods that demand strong, direct interaction with dissected tissue from the animal, either mechanically, such as patch clamp or multi-electrode arrays, or by intense laser light combined with the introduction of voltage-sensitive dyes or the expression of genetically encoded voltage imaging \cite{Emmenegger2019}. These procedures can be detrimental to the tissue under study due to the invasiveness of multiple electrodes or to the phototoxicity and cell toxicity of voltage sensitive dyes \cite{Emmenegger2019}. Results from experiments might depend on the setup and sample conditions such as the quality of electrical contact or localised optical properties of the surrounding tissue \cite{Okada1999} and can be specific to the biology of the target tissue, requiring adaptations between different tissues or structures. Finally, these techniques and in particular Ca$^{2+}$ imaging, can be unsuitable for measuring the axonal axial current, which is severely impaired by neurodegenerative diseases \cite{Compston2008}. 

A technique that allows passively, microscopic-scale recording of electrical activity in any kind of dissected living tissue without direct interaction is therefore desirable. One such method is to record the biomagnetic field induced by ionic currents associated with action potentials in neurons. Although such fields are extremely weak (nano- to femto-Tesla), magnetic field can freely permeate biological tissue with minimal interaction, avoiding damage and signal distortion and allowing remote recording. For low resolution sensing in whole human and animal subjects, magnetic field sensing has been implemented successfully as magnetoencephalography, giving centimetre-scale imaging of brain electrical activity from outside the body. However, this methodology relies on superconducting quantum interference devices (SQUIDs) \cite{Fagaly2006} or more recently spin-exchange relaxation-free (SERF) \cite{Boto2018} sensors, respectively demanding cryogenic cooling or a high temperature atomic vapour. Alongside other disadvantages, the intrinsically non-biocompatible properties of these sensors mean that these types of sensors cannot easily be brought into sufficiently close proximity (tens of micrometers) to the target sample necessary for microscopic-scale recording, particularly where a solution bath is required to keep tissue alive. What is therefore required is a new type of biocompatible sensor, capable of microscopically resolving and imaging electrical activity in dissected tissue \textit{in vitro} \cite{Barry2016}. In the past years, an alternative has emerged in the form of colour centres, optically active point defects in solid state materials \cite{Doherty2013}. Colour centres, in particular nitrogen-vacancy centres in diamond, have been employed as a new generation of quantum sensor \cite{Degen2017} to detect biomagnetic field induced by electrical activity \cite{Hall2012,Taylor2008}, offering remote and passive microsopic-scale recording of electrical activity based purely on fundamental physical principles, independent of the biological target system, in a highly biocompatible host material that can operate in high proximity and even within biological matter \cite{Kucsko2013}. 

In this work, we use a quantum sensor based on nitrogen-vacancy colour centres in diamond in a proof of principle experiment to demonstrate the first microscopic-scale recording of action potential propagation in tissue from the mammalian brain using biomagnetic field recording.  Our sensor operates via the principle of optically detected magnetic resonance (ODMR) spectroscopy \cite{Delaney2010}, where biomagnetic field modifies colour centre fluorescence emission under illumination with control fields (microwaves and laser light). We specifically target axial current associated with action potentials propagating in axons in the \textit{corpus callosum}, which connects the two cerebral hemispheres and enables integration of sensory-, motor- and cognitive information. This structure is very suitable for biomagnetic recording, having a high density of axons in a microscopic-scale region and a well-studied, strong and quantifiable biological response. Furthermore, the corpus callosum is of interest as small-scale functional changes and atrophy can be early indicators of neurodegenerative disease \cite{Ozturk2010, Pozzilli1991, Yamauchi1997,Teipel2002}. Using a tissue slice dissected from the brain of a mouse, with the slice kept alive \textit{in vitro} in a carbogenated solution bath, we record magnetic field from the action potentials propagating in callosal axons within a microscopic sensing region. Recording is performed without any direct interaction with the tissue, with the slice physically separated from the magnetic field sensor. We recover biomagnetic signals with distinct features that correspond to action potentials from both myelinated and unmyelinated callosal axons. We verify our biomagnetic recordings by simultaneous electrophysiological measurements using an invasive probe electrode inserted into the \textit{corpus callosum}. Finally, we demonstrate the capability to perform \textit{in situ} pharmacology on the tissue while recording, of importance for the study of target biosystem dynamics and for eventual evaluation of effectiveness of prospective drugs against disease. This is achieved by using a sodium channel blocker tetrodotoxin as an inhibitor, modifying action potential propagation \cite{DeCol2008}. 

\subsection{Biomagnetic recording of compound action potential}

\begin{figure}[!h]
\begin{centering}
\includegraphics[width=0.8\linewidth]{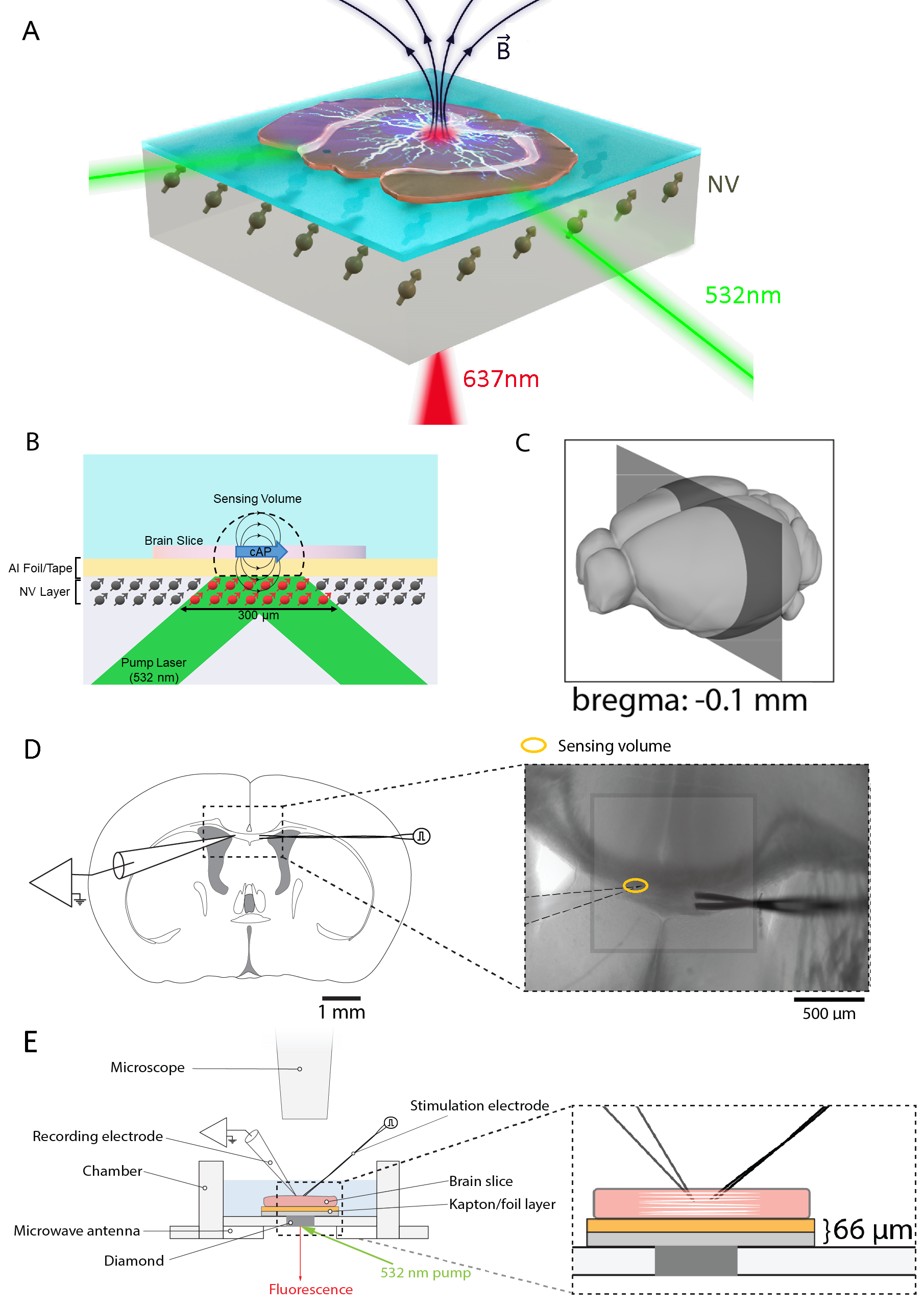}
\caption{\textbf{Sensor principle of operation, slice preparation and positioning and placement in the recording chamber} A) Schematic of the sensor operation (not to scale), where green laser light directed to subsurface colour centres in the diamond enables recording of magnetic field arising from compound action potentials (CAP) in a brain tissue slice placed above the diamond. B) Side-on schematic, detailing the microscopic sensing volume (dashed) above the NVs illuminated by the pump laser. C) 3D diagram of the mouse brain, showing the location of the slice used in experiments. D) Top-down illustration and image of a coronal brain slice from a mouse. The stimulation and invasive electrical recording electrodes are positioned in the \textit{corpus callosum} and the dark grey square indicates the position of the diamond beneath the slice, with the active sensing region marked. E) Side-on schematic of the recording chamber, with magnified central region. }
\label{Fig1}
\end{centering}
\end{figure} 

The operating principle of our sensor is illustrated schematically in Figure \ref{Fig1},A) with further details given in Methods. Using ODMR spectroscopy \cite{Barry2020}, magnetic field induced frequency shifts in the microwave spin resonances of the nitrogen-vacancy colour centres are transduced to changes in emitted fluorescence intensity $I_{FL}$ when pumped with 532nm laser light and low power microwaves. Changing fluorescence intensity $\Delta I_{FL}(t)$ can be recorded by a photosensitive detector and translated to a time dependent recording of absolute magnetic field strength observed at the diamond $\Delta I_{FL}(t)$ $\propto$ $\gamma_e$$B(t)$, using the gyromagnetic ratio of the nitrogen-vacancy electronic spin $\gamma_e$ = 2$\pi$$\times$28Hz/nT. Each spatially separate defect in a high density layer of centres close to the surface of the host diamond material can act as an independent sensor, allowing nanometer-scale resolution of magnetic field \cite{LeSage2013}. The colour centres are sub-surface, protected within the host material, yet close enough to the surface to accurately sense nearby magnetic field, making the sensor robust and biocompatible. They are also strong, stable optical emitters, which enables high magnetic field sensitivity with wide sensing bandwidth. Here we achieve sensitivity of 50pT/$\sqrt{Hz}$ with electronics-limited $f_{-3dB}$=10kHz using a 300$\times$100$\times$20$\mu$m$^3$ volume of colour centres addressed by the pump laser. We include a plot of the spectral response of our sensor in Supplementary Information. We highlight that the pump laser light is entirely contained within the diamond, and does not pass through or in any way directly interact with the solution bath or tissue under study. 

In Figure \ref{Fig1},B)  we show a schematic and white light image of our slice preparation, including electrode locations in the \textit{corpus callosum}. In C) we schematically show how the slice is introduced into our recording chamber, containing a bath of chilled, carbogenated artificial cerebrospinal fluid (ACSF) located above the sensor. The position of the slice could be freely manipulated in the chamber, with placement determined using a white light microscope mounted above, such that the \textit{corpus callosum} could be precisely located in the approximately 300$\times$100$\mu$m$^2$ sensing area above the diamond. The slice was separated from the diamond by an optically and electrically insulating layer, with sample-sensor distance approximately 60$\mu$m. Bath temperature above the diamond could be maintained at a stable 25C. Electrical activity in the \textit{corpus callosum} was induced by a bipolar stimulation electrode located 1-2 mm away from the sensing region, generating a compound action potential and associated ionic currents in callosal axons. For direct comparison to our biomagnetic recordings, compound action potentials were simultaneously recorded by measuring the local field potential with an invasive Pt/Ir electrode positioned on the contralateral side, closely adjacent to the biomagnetic sensing volume. We detail measurements from a limited number (n=3) of slices to provide proof-of-principle demonstration of our sensor; we do not seek in this work to perform statistical analysis of the observed biological response.

\begin{figure}[!h]
\begin{centering}
\includegraphics[width=154mm]{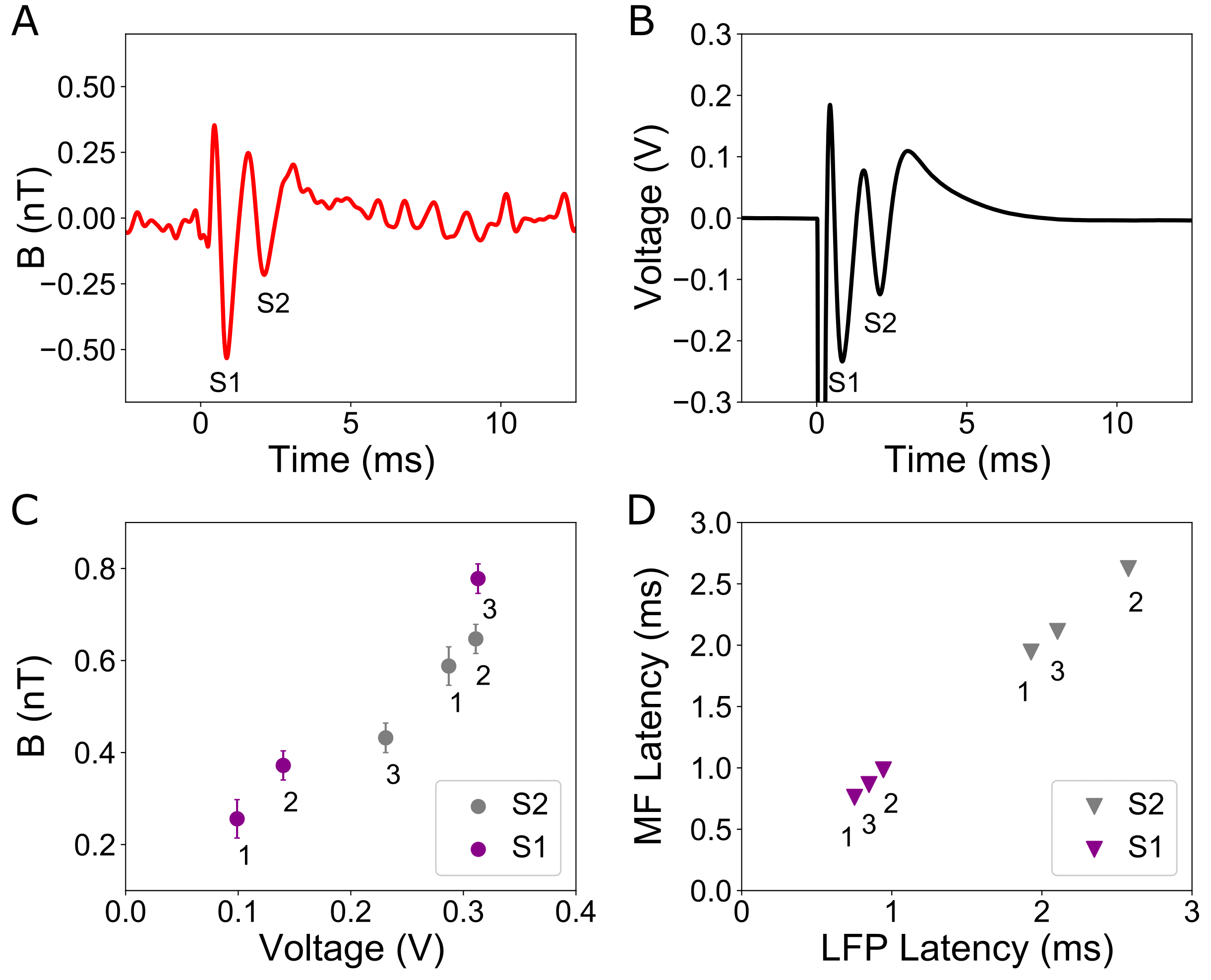}
\caption{\textbf{Detection of compound action potential from the \textit{corpus callosum}} A) Biomagnetic signal recorded by the quantum sensor from the \textit{corpus callosum}. B) Simultaneous local field potential (LFP) recording of the \textit{corpus callosum} compound action potential from the same region. Here, the tissue was stimulated electrically at $t$=0ms. Features S1 and S2 correspond to action potential from myelinated and unmyelinated axons respectively.  C) Amplitude of S1 and S2 as recorded by invasive electrical probe versus their amplitude from the biomagnetic recording. Here data is shown from 3 slices taken from different mice. The strength of the biomagnetic signal was linearly proportional to the amplitude of the local field potential. D) The same proportionality was also observed for latency, here shown for components S1 and S2 for invasive probe and biomagnetic recordings from the same three slices. }
\label{Fig2}
\end{centering}
\end{figure} 

We recorded a total of 28800 trials at a stimulation frequency of 2 Hz over a period of 4 hours from each individual slice. We recovered a detectable (SNR$>$5) biomagnetic signal when averaged over a minimum of 300 trials. Although single-trial readout would be preferable, our sensitivity is as yet insufficient to reach this level of performance. Figure \ref{Fig2},A) shows the simultaneously recovered electrophysiological signal of compound action potential propagation in the \textit{corpus callosum} and Figure \ref{Fig2},B) the biomagnetic signal, recorded and averaged over 60 minutes. We recovered a biomagnetic signal from our quantum sensor with the same characteristic features as the local field potential signal recorded electrically by the invasive probe. Our signal arises from basic physical principles for the magnetic field induced by a flow of electrical current (Ampere's circuital law). We observed two distinct features in both datasets, labelled S1 and S2 in Figure \ref{Fig2}. Features S1 and S2 are understood in the literature to arise from compound action propagation primarily in myelinated and unmyelinated axons respectively \cite{Crawford2009}, with some continuum of overlap (i.e. a minor myelinated contribution to S2 and \textit{vice versa}). The relative amplitude and latency characteristics were the same for both S1 and S2 in both biomagnetic and electrical probe data. S1 and S2 in the biomagnetic readout changed in amplitude and latency in linear proportion to the equivalent local field potential, with a stronger electrical response corresponding to a stronger biomagnetic field signal. This proportionality was consistent when the experiment was repeated for multiple slices from different mice, as shown in Figure \ref{Fig2},C) and D). We note that the ratio between amplitudes of S1 and S2 can vary depending on the ratio of myelinated to unmyelinated axons in the corpus callosum, which can vary significantly between individual animals \cite{Reeves2012}. An advantage of biomagnetic recording was the minimisation of the strong stimulation artifact electrically induced in the local field potential recording, offering the benefit of recovery of signal features much closer to the stimulation time ($t$$<$0.5ms). Full details of how this was achieved is detailed further in Supplementary Information. 

\subsection{\textit{In-situ} pharmacology with tetrodotoxin}

\begin{figure}[!h]
\begin{centering}
\includegraphics[width=0.8\textwidth]{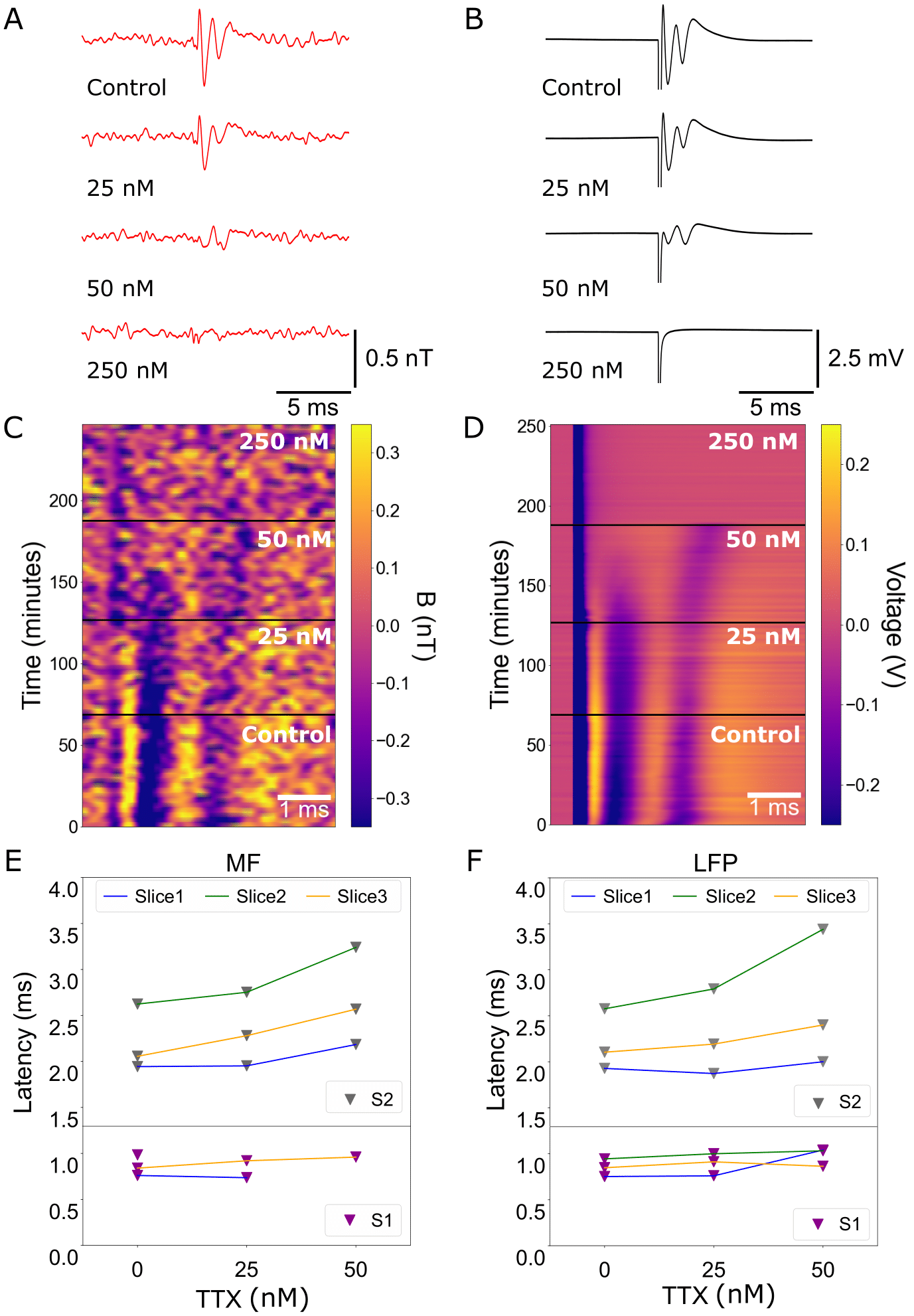}
\caption{\textbf{The effect of the introduction of tetrodotoxin on compound action potential propagation in the \textit{corpus callosum}} A) Local field potential (LFP, black) measure by invasive electrical probe and simultaneous biomagnetic signal from the quantum sensor (red) recorded from the \textit{corpus callosum} as a function of increase in tetrodotoxin concentration over the 4 hour duration of the experiment. B), C) Color plots of the LFP and quantum sensor recording as the experiment progressed. Each y-axis step represents the average of 360 trials (3 minutes of recording). In B and C, S1 and S2 components appear in purple. D), E) Quantification of the latency shift in the LFP and biomagnetic field signal respectively caused by tetrodotoxin for S1 (bottom) and S2 (top) for 3 tissue slices from different mice.}
\label{Fig3}
\end{centering}
\end{figure}

To determine whether our sensor could detect subtle changes in axon action potential waveform, comparable to what happens in the course of neurodegenerative diseases \cite{Hamada2017}, and to illustrate our capability to perform simultaneous pharmacology during recording, we inhibited the action potential signal by blocking voltage sensitive sodium ion channels using tetrodotoxin \cite{Narahashi1964}. This was added in increasing nanomolar concentration to the solution bath containing the brain tissue slice while recording from the \textit{corpus callosum} using the quantum sensor and again simultaneously with an invasive electrical probe. The resulting effect of increasing tetrodotoxin concentration on the electrical signal can be seen in Figure \ref{Fig3}, A) and C) with the comparative effect on the biomagnetic signal in Figure \ref{Fig3}, B) and D). After a 60 minute control recording, concentration was increased to 25nM, then to 50nM after a subsequent 60min recording period, with final large concentration of 250nM added after 3 hours of recording, to fully inhibit the signal. We include in Supplementary Information the equivalent data plots for a control slice with no introduction of tetrodotoxin.

In the magnetic data recorded by our quantum sensor and the simultaneous electrical recording, we observed the reduction in amplitude and elimination of the fast S1 signal feature and the increase in latency of the slower S2 feature, up to and including concentrations of 50nM. The reduction in S1 and S2 amplitude has been previously established to arise through decrease in the number of functional sodium ion channels. Myelinated axons have a high proportion of more TTX-sensitive Nav$_{1.6}$ sodium channels compared to unmyelinated axons which mainly express the less-TTX sensitive Nav$_{1.2}$ sodium channels, making myelinated axons more susceptible to tetrodotoxin \cite{Caldwell2000,Yamano2022,Catterall2005}. This was specifically observed as a faster decrease in amplitude of feature S1 with increasing levels of tetrodotoxin compared to S2. The blocking of sodium channels was also observed as a decrease in the conduction velocity of unmyelinated axons, seen as an increase in the latency of signal S2 with higher concentration \cite{DeCol2008}. These behaviours were again observed consistently in both magnetic and electric recordings and between tissue slices from different mice as shown in Figure \ref{Fig3},D) and E). In the presence of high tetrodotoxin concentration (250 nM), the biological signal was abolished by blocking the majority of sodium ion channels in both myelinated and unmyelinated axons, in order to verify our recording of action potential rather than artifact signal.  

In summary, we have successfully demonstrated the recording of neural activity from living brain tissue at the microscopic scale using a diamond quantum sensor. As our technique directly records magnetic field, no modification of the target system biology is required, and therefore the technique is able to record ionic current from any region within any type of tissue section. The technique is sensitive enough and has sufficient time resolution to detect subtle changes in electrical activity. This we demonstrate in a specific target region of the brain, axons in the \textit{corpus callosum}, recording the biomagnetic field from action potentials arising from axons of different types (myelinated and unmyelinated), with good agreement with simultaneous control measurements using an electrical probe.  Our technique can enable detailed microscopic study of pathologies affecting neuronal electrical activity in animal disease models. Although our work represents a far earlier stage of technological development, it offers an eventual route to advantages over well-established microscopic techniques. This includes entirely passive recording with minimal direct sample interaction and no need to modify the target biological system, free of the disadvantages of alternative techniques (probe arrays, Ca$^{2+}$ imaging, genetically encoded voltage inducators) such as distortion by the relative conductivity surrounding tissue, the need for mechanical contact or direct laser illumination of the sample, reliance on the presence of specific ion types or the need for biological modification of the target tissue through genetic modification or dye introduction. Finally, the technique offers a route to high spatial imaging resolution by exploiting the high, consistent densities of colour centres, far greater than in-tissue fluorescence markers or existing microfabricated probe arrays \cite{2107.14156, Levine2019}. 

\clearpage

\section{Methods}

\subsection{Diamond Quantum Sensor}
Our sensor is based on a [100] oriented electronic-grade diamond crystal (Element Six). Diamond dimensions were 2 $\times$ 2 $\times$ 0.5 mm$^3$. The diamond was overgrown via chemical vapour deposition (CVD) to a thickness of approximately 20$\mu$m thick diamond layer with $\approx$5 ppm of $^{14}$N doping in the overgrowth layer. N to NV conversion was achieved by proton irradiation at 2.8MeV followed by annealing at 800\textdegree C in an inert atmosphere. Our ODMR resonance linewidth was 1 MHz and contrast of 1.5\% for a single NV axis. For dissipation of heat generated by the pump laser, we mounted the diamond in a laser cut aluminium nitride heatsink. The top surface of the diamond was covered by 16$\mu$m thick aluminium foil, acting as an additional heatsink and reflecting pump laser back into the diamond. A layer of Kapton tape (50$\mu$m) above the foil electrically insulated the sample from the diamond. Microwave were supplied by a custom-designed printed board microwave antenna placed below the diamond and AlN heatsink. Our biological tissue was contained in a solution bath in a 3D printed plastic chamber, mounted directly above the diamond and made watertight using bio-safe silicone.

We optically pumped our NV centres with 1.2-1.4W of linearly polarised single mode 532nm green laser light (Coherent Verdi G2), coupled into the diamond at Brewster's angle (67\textdegree) from beneath the diamond. No pump light passed through the solution bath or biological sample. NV centre fluorescence was collected by a 12mm condenser lens (Thorlabs ACL 1210) and the pump light removed by an optical longpass filter (FEL0600). Biomagnetic field was recorded via changes in intensity of red fluorescence emission using an electronically balanced optical receiver (Nirvana 2007, New Focus Inc.). We measured magnetic field response only in a single direction, defined by a static bias field of 1.5 mT applied parallel to the diamond [110] crystallographic direction. We used a continuous wave scheme with three-frequency microwave driving \cite{ElElla2017} using two microwave generators (Stanford SG394) tuned to the spin transition (2.7-3GHz) and the $^{14}$N hyperfine transitions (2.16 MHz). The microwaves were modulated at 33.3 kHz for lock-in detection (Stanford SR850), with time constant of 10$\mu$s leading to a -3dB rolloff measurement bandwidth of $\approx$10kHz. Lock-in amplifier output was digitised at 125 kSa/s by an analog to digital converter (ADC, NI PCI-6221) and processed using custom Labview software. 

Data was simultaneously acquired using a Pt/Ir electrode probe, inserted into the corpus callosum in close proximity to the colour centre sensor. Readout was performed using a preamplifier stage and amplifier  (Axon Instruments) and recorded by the same ADC as the magnetic data. To ensure no crosstalk between the electrical probe and the magnetic data, the amplified output voltage was minimised and decoupled from the colour centre sensor measurement electronics using an optocoupler. Tests were performed with the colour centre sensor microwave drive frequency set away from the NV centre resonance frequency (magnetically insensitive) while recording electrically from the slice; no crosstalk signal was observed in the magnetic data.

\subsection{Brain Slice Preparation}

Adult mice (5-10 weeks) were anaesthetized with isoflurane, decapitated and brains were rapidly dissected submerged in ice-cold, carbogen (95\% O$_2$ / 5\% CO$_2$) saturated sucrose substituted artificial cerebrospinal fluid (ACSF) containing (in mM): 200 Sucrose, 11 Glucose, 25 NaHCO$_3$, 2.5 KCl, 0.5 L-ascorbic acid, 2 Na-pyruvate, 3 myo-inositol, 1.25 NaH$_2$PO$_4$, 0.1 CaCl$_2$, 4 MgCl$_2$. 400 $\mu$m thick coronal slices containing the \textit{corpus callosum} were obtained using a Leica VT1200s vibratome (Leica Biosystems, Germany). Slices were transferred to an interface holding chamber filled with carbogen saturated regular ACSF containing (in mM): 111 NaCl, 11 Glucose, 25 NaHCO$_3$, 3 KCl, 1.1 KH$_2$PO$_4$, 2.5 CaCl$_2$, 1.3 MgCl$_2$ and allowed to recover at 28C for at least 1 hour. The slices were introduced into the same carbogenated ACSF solution, continuously circulated from a reservoir into the solution bath. We chilled the inflow of the ACSF solution into the bath by passing the feed line through an ice bath. This helped dissipate heat from the diamond and AlN heatsink plate, allowing us to maintain the region directly above the sensor at a stable 25C for up to 24 continuous hours. The slice was held down onto the sensor using nylon strings on a nonmagnetic Pt harp structure, ensuring fixed close proximity to the diamond. Our method is not limited to the corpus callosum and allows precision placement of any brain region within the sensing area. The corpus callosum was located using top-down white light microscopy, recording video through a digital camera (IDS) at 60 frames-per-second while positioning the slice. 
 
\subsection{Data Processing}

Measurements were performed without magnetic shielding in an ordinary laboratory (basement) environment. We removed background magnetic noise from electrical mains (primarily 50,150Hz) and non-stationary sources (nearby heating and water pumps) by frequency domain notch filtering, using methodology outlined in detail in our previous works \cite{Webb2021, Webb2020}.  Filtering was performed in a three stage process. First, to avoid ringing,  the artifact from the fast stimulation spike (50 $\mu$s) was fitted in the time domain and subtracted (full methodology detailed in Supplementary Information). Secondly, the data was fast Fourier transformed to the frequency domain and the magnetic noise at e.g. 50/150Hz was removed by notch filters. As our signal SNR$<<$1 for a single 60sec time trace, the notch frequencies were identified by a threshold factor of 2-3$\times$ above the white noise floor. A lowpass filter with upper cutoff at 2.5 kHz was implemented to restrict the sensing bandwidth to the frequency range of the target biosignal. In the third and final stage, the signal was inverse Fourier transformed back to the time domain. Epochs of 100 ms, centered on the stimulation triggers were extracted from the filtered data, for a total of 120 in each 60sec acquisition. These epochs were extracted and averaged to reveal the desired biological signal.



\subsection{Funding}
This work was funded by the Novo Nordisk foundation through the synergy grant bioQ (Grant Number: NNF17OC0028086) and the Center for Macroscale Quantum States (bigQ) funded by the Danish National Research Foundation (Grant number:DNRF142). Hartwig R. Siebner holds a 5-year professorship in precision medicine at the Faculty of Health Sciences and Medicine, University of Copenhagen funded by the Lundbeck Foundation (Grant Nr. R186) while Kirstine Berg-Sørensen acknowledges support by Independent Research Fund Denmark (grant no 0135-00142B) and Novo Nordisk Foundation (grant no NNF20OC0061673).


\subsection{Ethics declarations}
All methods in this work were carried out in compliance with the ARRIVE guidelines according to relevant Danish national guidelines and regulations. Experimental protocols were approved where required by the Technical University of Denmark, the University of Copenhagen and the Danish National Committee on Health Research Ethics (DNVK). 

\subsection{Authors' contributions}
!REF to be added

\bibliographystyle{unsrt}

\end{document}